\documentclass[manuscript, screen]{acmart}

\usepackage{natbib}

\AtBeginDocument{%
  \providecommand\BibTeX{{%
    Bib\TeX}}}

\setcopyright{acmcopyright}
\copyrightyear{2022}
\acmYear{2022}
\acmDOI{XXXXXXX.XXXXXXX}

\acmJournal{JACM}
\acmVolume{37}
\acmNumber{4}
\acmArticle{111}
\acmMonth{8}

\begin{document}

\title{\textit{E Pluribus Unum}: \\Guidelines on Multi-Objective Evaluation of Recommender Systems}

\author{Patrick John Chia}
\affiliation{%
  \institution{Coveo}
  \city{Montreal}
  \state{Quebec}
  \country{Canada}
}
\email{pchia@coveo.com}

\author{Giuseppe Attanasio}
\affiliation{%
  \institution{Bocconi University}
  \city{Milan}
  \country{Italy}
}

\author{Jacopo Tagliabue}
\affiliation{%
  \institution{Bauplan}
  \city{New York}
  \state{NY}
  \country{USA}
}
\affiliation{%
  \institution{NYU}
  \city{New York}
  \state{NY}
  \country{USA}
}

\author{Federico Bianchi}
\affiliation{%
  \institution{Stanford University}
  \city{Palo Alto}
  \state{CA}
  \country{USA}
}

\author{Ciro Greco}
\affiliation{%
  \institution{Bauplan}
  \city{New York}
  \state{NY}
  \country{USA}
}
\author{Gabriel de Souza P. Moreira}
\affiliation{%
  \institution{NVIDIA}
  \city{Sao Paulo}
  \country{Brazil}
}
\author{Davide Eynard}
\affiliation{%
  \institution{mozilla.ai}
  \city{London}
  \country{United Kingdom}
}
\author{Fahd Husain}
\affiliation{%
  \institution{mozilla.ai}
  \city{Toronto}
  \country{Canada}
}

\renewcommand{\shortauthors}{Chia et al.}

\begin{abstract}

Recommender Systems today are still mostly evaluated in terms of accuracy, with other aspects beyond the immediate relevance of recommendations, such as diversity, long-term user retention and fairness, often taking a back seat. Moreover, reconciling multiple performance perspectives is by definition indeterminate, presenting a stumbling block to those in the pursuit of rounded evaluation of Recommender Systems. EvalRS 2022 --- a data challenge designed around Multi-Objective Evaluation --- was a first practical endeavour, providing many insights into the requirements and challenges of balancing multiple objectives in evaluation. In this work, we reflect on EvalRS 2022 and expound upon crucial learnings to formulate a first-principles approach toward Multi-Objective model selection, and outline a set of guidelines for carrying out a Multi-Objective Evaluation challenge, with potential applicability to the problem of rounded evaluation of competing models in real-world deployments.

\end{abstract}



\received{20 February 2007}
\received[revised]{12 March 2009}
\received[accepted]{5 June 2009}

\maketitle

\section{Introduction}

Recommender Systems (RSs) can be evaluated according to different quality factors, like relevance/accuracy, diversity, novelty, serendipity, coverage, and fairness, among others \cite{zangerle2022evaluating}. However, recommendation accuracy, conceived as \textit{immediate item relevance} --- ``Is the user going to watch the recommended movie?''
\cite{bennett2007netflix} or ``Is the user going to buy the suggested product'' \cite{Virinchi2022,areUSure} --- has been by far the main measure to assess the performance of RSs.

Industry practitioners and academics are becoming increasingly aware that a focus on only \textit{accuracy} as a performance objective can result in RSs that transfer poorly ``in-the-wild''. While improving business metrics, accuracy-oriented systems have been shown to have unintended, yet profound, social ramifications \cite{YoutubeBadRecs, 10.1145/3292500.3330691,raj2022fire,10.1145/3523227.3551480} such as inciting social divisiveness \cite{FacebookBadRecs} or spreading misinformation \cite{youtuberegret}. This behavioural gap calls for a re-assessment of \textit{how} we evaluate RSs, shifting from a narrow focus on singular accuracy metrics to a more rounded evaluation methodology that considers a wider range of metrics.

Numerous works have introduced new RS quality measures \cite{zangerle2022evaluating}, such as fairness, diversity, and serendipity, of which, some have even been used in research focusing on multi-objective training of RSs \cite{10.1145/3298689.3346998}, e.g., optimizing for both accuracy \textit{and} diversity \cite{adomavicius2011improving}.
Be that as it may, relatively little effort has been put into designing and operationalizing \textit{multi-objective evaluation}, i.e., ranking models in light of multiple quality metrics, in part due to factors such as the indeterminate nature of aggregating multiple metrics, difficulty in modeling the relationship between them, and combinatorial explosion. For example, given an accuracy metric such as Normalized Discounted Cumulative Gain (\texttt{NDCG}) and a fairness metric such  Popularity Bias (\texttt{PB}), it is not known \textit{a priori} whether they are correlated, what the relationship (e.g., linear, exponential, quadratic) between them is, or how they should be combined (e.g., a weighted sum, ratio). Moreover, the problem quickly becomes more difficult when more metrics are considered.

In this paper, we reflect on our experience of organizing EvalRS 2022 \cite{evalrs2022}, a first-of-its-kind data challenge that focused on the rounded evaluation\footnote{``Rounded evaluation'' was the original description of EvalRS, while ``multi-objective evaluation'' is a more formal framing for the challenges we tackle in this paper: we will use both phrases interchangeably, as there is no risk of confusion.} of RSs, hosted at the CIKM 2022 AnalytiCup, and draw upon it to introduce a first-principles approach toward multi-objective evaluation. We provide a holistic analysis of the challenge, encompassing our past experience and motivations, present reflections, and future possibilities. While the motivation for this work comes directly from a data challenge, we believe many of our learnings may generalize to the real-world, and have broad applicability in deployment scenarios when competing models needs to be evaluated.
Section \ref{sec:background} provides some background on multi-objective evaluation.
Then, in Section \ref{sec:evalrs2022}, we review our past experience provided by EvalRS 2022. Next, in Section \ref{sec:rounded_eval}, we reflect on the shortcomings and limitations of EvalRS 2022, where we propose, describe, and implement a more principled approach toward evaluating models in light of multiple metrics. Lastly, in Section \ref{sec:guidelines}, we summarize our findings into guidelines for future challenges. 

\section{Background}
\label{sec:background}

In this section, we will introduce the formalism and main components that are required to understand our work, and to situate our contributions in the appropriate context.

\subsection{Beyond-Accuracy Metrics}

Standard evaluation of RS tend to revolve around accuracy, i.e., how well a system's recommendations match implicit or explicit user feedback. Recent advances have focused instead on novel quality characteristics, such as diversity (dissimilarity of recommendations under certain criteria) or coverage (the number of items a system can or will recommend), serendipity, novelty, fairness across users or groups, and more (see Table 5 in \cite{zangerle2022evaluating}). We refer to such metrics as \textit{beyond-accuracy metrics}. 

A key focus of EvalRS 2022 was group fairness. One practical way of measuring group fairness is by choosing a utility function and measuring how it varies across subgroups (e.g., demographics \cite{dixon2018measuring}). Miss-Rate Equality Difference (\texttt{MRED}) \cite{evalrs2022} requires computing the miss-rate ($MR$) for each group of interest and measuring how far it deviates from the global miss-rate:

\begin{equation}
    \texttt{MRED} = -\sum_i{ | MR_i - MR_{avg} | } \; \;  \forall{i \in G}
\end{equation}

Where $MR_{avg}$ is defined as the MR computed on the entire test set, $MR_i$ is defined as the MR computed on a group $i$, and $G$ is a partitioning of the data into groups under a given criterion (e.g., partitioning by Country).


\subsection{Multi-Objective Evaluation}
\label{ssec:background_moe}

\begin{figure}[h!]
\centering
\includegraphics[width=0.55\linewidth]{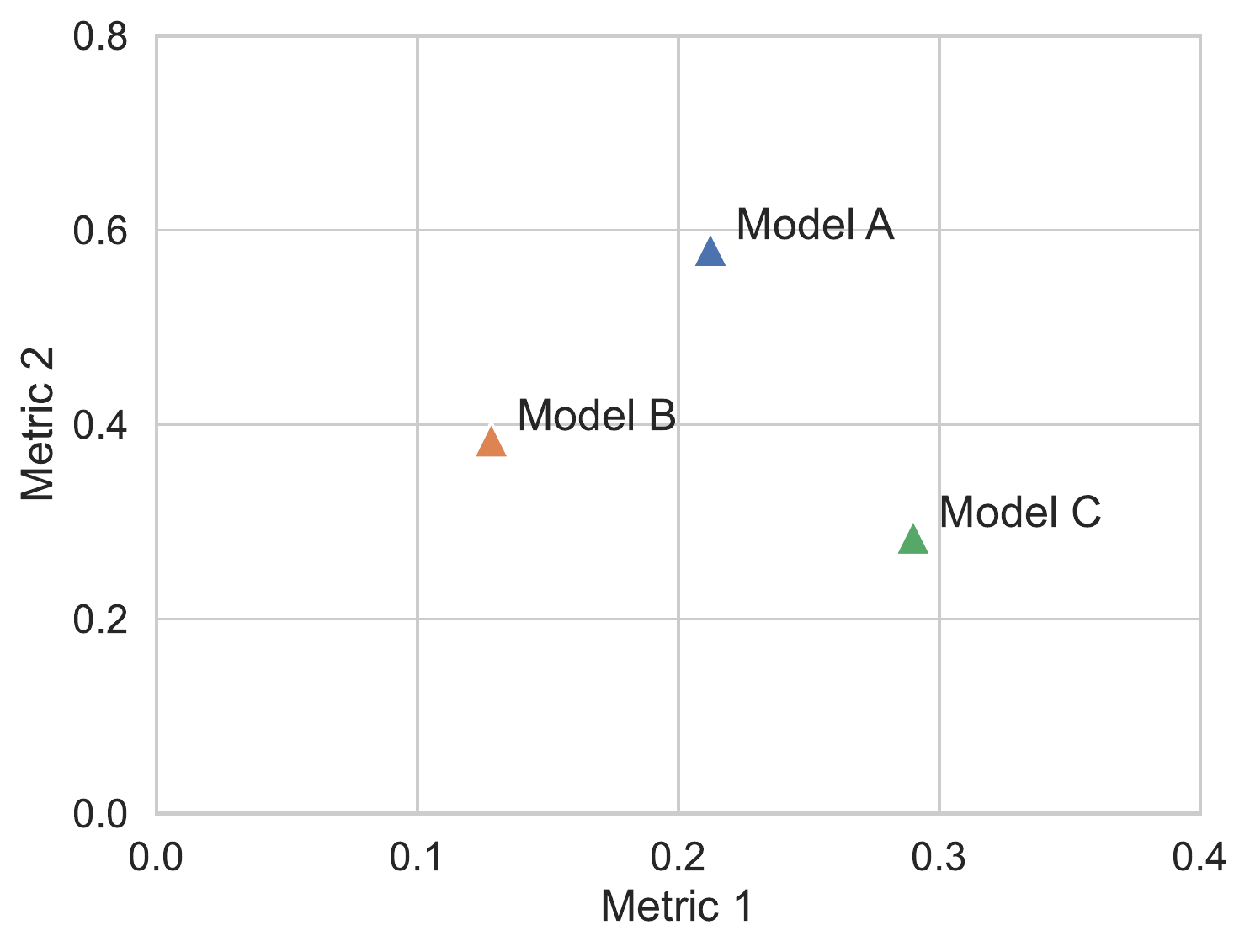}
\caption{Illustration of the concepts of \textit{pareto-dominance} and \textit{pareto-noncomparability}. Pareto-Dominate: Model A performs better than Model B in both Metric 1 and Metric 2. Pareto Non Comparable: Model A performs better than Model C in Metric 2 but Model C performs better than Model A in Metric 1.}
\label{fig:pareto_dominanace}
\end{figure}

Multi-Objective Evaluation refers to the process of ranking models based on two or more metrics/objectives, in order to select the ``best'' performing model.  Considering the simplest multi-objective case of two dimensions in Fig. ~\ref{fig:pareto_dominanace}, whereby the goal is to maximize both Metric 1 and Metric 2, we see that while it is clear that Model A is better than Model B, it is not immediately clear whether Model A is better than Model C. Borrowing from the literature of Multi-Objective Optimization \cite{ehrgott_2000}, we can say that Model A \textit{pareto-dominates} Model B, while Model A and Model C are \textit{pareto-noncomparable}. In fact, the space of nondominated models, which is a subset of noncomparable models, grows exponentially with the number of dimensions involved \cite{ALLMENDINGER2022105857}. As such, without making any further assumptions, when multiple objectives are in consideration there is no well defined meaning of a ``best'' model, i.e., pareto-nomcomparable models are equally ``good''.

Rounded evaluation thus requires one to devise a scoring function $S(.)$ which takes in $k$ metrics $<m_0, ..., m_{k-1}>$ from a Model $M$ and returns a single score $s \in \mathbb{R}$ which can be used for model ranking and selection. It is worth noting that while \textit{prima facie} this might have a similar appearance as a single-objective metric, the underlying mechanism is multi-faceted and hence different: how $S(.)$ is defined will ultimately determine its utility in rounded evaluation -- the goal is not simply to ``come up with a single number'' \footnote{Or, we could have just pick the scoring function $S(m_0, ..., m_{k-1}) = m_0$, and obtain a single-dimensional metric we are familiar with.}, but to pick a number that conveys a (for the lack of a better word) rounded view of model behavior. Careful thought must therefore be put into the various design decisions that go into the construction of $S$.

\subsection{Multi-Objective Evaluation vs. Multi-Objective Optimization}

The interplay between multi-objective optimization and multi-objective evaluation is subtle. On the one hand, multi-objective optimization is concerned with \textit{searching} for models which perform well under a set of objectives, where common challenges involve, but are not limited to, operationalizing multiple metrics as optimizable objectives (e.g. differentiable), handling non-convex objectives, and finding solutions (models) that are pareto optimal.
On the other hand, multi-objective evaluation deals with the ranking of a given set of models when multiple measures of performance are involved, i.e., ``how do we make sense of multiple metrics and ultimately select the best candidate?''. While both appear to deal with some form of model selection in the context of many objectives, multi-objective evaluation often takes things a step further: multi-objective optimization is generally interested in finding a set of models which are optimal in the pareto-optimal sense, whereas, multi-objective evaluation is interested in picking a single model based on certain assumptions (e.g. accuracy is more important that fairness). That being said, the interdependence of both perspectives is undeniable, too, and techniques and ideas from the multi-objective optimization domain often have analogs and can offer insight into multi-objective evaluation.

\section{The Past: EvalRS 2022, Multi-Objective Evaluation in Practice}
\label{sec:evalrs2022}
In light of the current limitations in accuracy-centric testing, we organized EvalRS 2022 as a first-of-its-kind data challenge to foster discussion and innovation around rounded evaluation for RSs, with the aim of bridging the gap between offline evaluation and performance ``in the wild'' \cite{tagliabue2023challenge}. 
In this section, we introduce the key components of EvalRS 2022 \footnote{EvalRS 2022, the first edition of the challenge and workshop, counted more than 150 participants, divided into 50 teams from 14 countries, distributed across industry and academia. Please find more details at \url{https://github.com/RecList/evalRS-CIKM-2022}.} and several core guiding principles and learnings from our first attempt at a practical implementation of multi-objective evaluation. 

\subsection{Challenge Overview}

EvalRS 2022 presented participants with a classical user-item recommendation task in the music domain: the goal is to recommend each user an appropriate song given their listening history and a catalog of target items. For the task, we utilized \textit{LFM-1b} \cite{10.1145/2911996.2912004}, a large-scale dataset with information on music consumption on \textit{Last.fm} and performed preprocessing (summary statistics of the final dataset in Table \ref{table:dataset_stats}) such that it was accessible to, and amenable for, a wide range of participants and compute environments.

\begin{table}[!t]
\centering
\caption{\label{table:dataset_stats} Descriptive statistics for the EvalRS dataset.}
\begin{tabular}{lr} 

 \toprule
 Items & Value \\
 \midrule
 Users & $119,555$ \\ 
 Artists & $62,943$ \\ 
 Albums & $1,374,121$ \\ 
 Tracks & $820,998$ \\ 
 Listening Events & $37,926,429$\\ 
 User-Track History Length (25/50/75 pct) & $241/346/413$ \\ 
 \bottomrule
\end{tabular}
\end{table}

\subsection{Initial Guidelines}

We designed EvalRS 2022 around an initial set of guiding principles: we share those guidelines and the reasoning behind them and highlight the strengths and weaknesses of our original design, which motivated this follow-up contribution.


\subsubsection{Adopt Diverse Evaluation Metrics.}
\label{ssec:evalrs_scoring}

Our primary goal of EvalRS 2022 was to promote the development of models that balance between standard operating requirements, e.g., retrieval accuracy, \textit{and} more uncommon (yet important) ones, e.g., being fair across subgroups or demonstrating robustness in recommendation. Therefore, following \cite{DBLP:journals/corr/abs-2111-09963}, we proposed to evaluate models against a comprehensive ``reclist'' which included both accuracy and beyond-accuracy metrics. Specifically, on top of the usual retrieval metrics such as Hit-Rate (HR) and Mean Reciprocal Ranking (MRR), we introduced beyond-accuracy metrics in the form of behavioral and fairness tests.
 
Behavioral tests included a ``being-less-wrong'' \cite{DBLP:journals/corr/abs-2111-09963} test --- measuring whether a wrong recommendation remains relevant (e.g., if it has the same genre of the goal recommendation) --- and diversity. Both tests leveraged a latent space of tracks learnt from historical user activity data.  

 The fairness tests were focused on group fairness: using Miss Rate (MR) as the utility metric, we tested for utility equality across groups with MRED \cite{evalrs2022}. We extracted groups by slicing across gender, country, listening activity, and track popularity and artist popularity. 

Additionally, as part of the EvalRS challenge, we asked participants to come up with a novel beyond-accuracy metric to encourage new evaluation perspectives.

\subsubsection{Rigorous Evaluation Protocol}

The reproducibility crisis in RS research and evaluation is a known problem \cite{10.1145/3434185, 10.1145/3298689.3347058, sun2020we}. Following the recommendations of \citet{schnabel2022we}, we thus adopted Bootstrapped Nested Cross Validation (BNCV) as the evaluation protocol of EvalRS 2022. The protocol performs takes a learning algorithm (i.e., a procedure that trains and returns a recommendation model) and performs $N$ iterations of training and evaluation on $N$ random samples (with replacement) of the entire dataset. Sampling was conducted by stratifying on users. The model is then assigned its average performance, per-metric, across iterations.

BNCV fulfilled two practical needs. From the perspective of a data challenge, evaluation across multiple folds of the data prevents participants from overfitting on the dataset\footnote{LFM-1b is a publicly available dataset. A single train/test split approach would too easily allow for test set leaking.}. From a model evaluation perspective, it minimizes the chance of ``false discoveries,'' i.e., performance gains simply due to a lucky split, thereby increasing the overall reliability of our evaluation approach.

\subsubsection{A Two-staged Evaluation}

Recent evidence has shown that recommendation quality measures such as diversity or fairness either compete with or favor accuracy differently \cite{jannach2022multi}. However, such relationships are not generally known in advance, especially in the context of a data challenge. We thus proposed a two-stage evaluation, each with different leaderboards and scoring strategies, allowing us to leverage submissions from the former stage to make a more informed model score in the latter. In each stage, we summarize all tests into a single scalar score.

During the first stage, the scores from each individual test were simply averaged; here, tests have equal importance. While getting participants acquainted, this stage has the crucial goal of gathering data on the relative difficulty of tests -- the average itself does not matter: participants have granular access to single test results to build intuition on what the model does well (or not), and they are actively discouraged from ``interpreting'' the aggregate score.
At the end of this phase, we might notice, for example, a test where, overall, models underperformed relative to other tests (e.g., MRED on Gender vs. MRED on other groups). We would hence assume it to be hard under the current challenge setup.

Leveraging data collected from stage one, the second stage applied a different scoring mechanism. Here, we apply a weighted average where each test contribution is given by how a submission placed compared to lower and upper baselines. For a given metric $k$, each score was hence transformed as:
\begin{equation}
    \label{eqn:evalrs_norm}
    \hat{m_k} = \frac{m_k - m_k^{base}}{m_k^{best} - m^{base}}
\end{equation}
where $m_k$ is the score for  metric $k$, $m_k^{base}$ is the score for $k$ of a baseline we provided, and $m_k^{best}$ is the score for $k$ of the best stage one submission.

As such, each metric is recalculated based on how much it out performs the baseline ($m_k - m_k^{base}$) relative to how much the best submission out performs the baseline ($m_k^{best} - m^{base}$), thereby  normalizing scores across metrics. After the normalization, we used a weighted mean to compute the final score, giving slightly more weight to fairness and behavioral categories over accuracy. See \citet{evalrs2022} for further details.

Despite our best efforts at designing a fair model scoring methodology, we believe this to be the key weakness of EvalRS 2022, and that in retrospect, the normalization procedure used could be improved significantly. In particular, we observed that (1) the normalization procedure only utilizes information from the baseline and best model and (2) it hinges upon our choice of best model and baseline, with an overall result of confounding metric aggregation and metric importance weighting. In Section~\ref{sec:rounded_eval}, we introduce a more principled approach toward model scoring that takes full advantage of the data collected in the first stage.

\subsubsection{Accessible Evaluation APIs}

We believe that the theoretical discussion on multi-objective evaluation should come along with an easy-to-access, reproducible, and extensible programmatic evaluation API. Bringing together the above guidelines and principles, we released, as part of EvalRS 2022, open-source training and evaluation code along with resources (e.g., tutorials, blogs, notebooks) and strong baselines. Our evaluation framework builds around RecList \cite{DBLP:journals/corr/abs-2111-09963}, an open-source Python library to automate behavioral testing of RSs. We implemented as well, the score normalization and Bootstrapped Cross-Validation procedures described above into a single abstraction. As a result, participants needed only to provide code for model training. Since everything is available as open source code and the evaluation can be performed locally without any organization or server available, EvalRS 2022 dataset and tests will remain a forever accessible benchmark to test models on, even long after the initial competition ended. 

\subsection{Best Contributions to EvalRS 2022}
\label{sec:evalrs_workshop}

The participating teams of EvalRS developed original and effective systems. Here, we highlight the best solution proposed by the winning team and the best beyond-accuracy metric introduced.

\subsubsection{Best Solution}
\citet{park2022item} proposed an ensemble between Variational Auto-Encoders (VAEs) and Bayesian Personalized Ranking Matrix Factorization. The proposed VAE architecture is structured following an Item-oriented formulation (rather than the most typical User-oriented VAE formulation). The authors noted that Item-oriented VAEs tend to recommend less popular items. Further, they introduced a fairness regularization term in the loss, following \citet{borges2022f2vae}. However, the authors observed that while Item-VAEs, had better fairness across item popularity, it suffered in performance in the ``being less wrong'' test.

\subsubsection{Best Test}
The Variance Agreement test introduced in ~\citet{giobergia2022triplet} was selected as the best test. This metric measures how well the diversity in a recommendation matches the user's tendency to prefer diverse items (i.e. a certain user may listen mostly to a single genre of music, whereas another might prefer a broader range of genres). The author computes the diversity using the Gini impurity coefficient. Notably, the test applies to any user preference: in the challenge, the authors tested diversity in terms of different artists recommended.

\section{The Present: How to Perform a Rounded Evaluation}
\label{sec:rounded_eval}

The main weakness of EvalRS 2022 was in its scoring methodology. We, therefore, propose a more principled approach toward rounded evaluation, one that is derived from first principles and one which draws upon the challenges and experiences we faced when organizing the first edition of EvalRS.

We begin by illustrating \textit{why} rounded evaluation is difficult, then we propose \textit{how} one can systematically reason about the problem, and offer a possible solution. Lastly, we demonstrate the merits of our newly proposed evaluation approach and provide further intuition by comparing it with the methodology used in EvalRS 2022.

\subsection{The Challenge of Rounded Evaluation}
\label{sec:eval_chall}

As remarked in \S \ref{ssec:background_moe}, multiple metrics make models comparable across interesting dimensions, at the cost of making it harder to \textit{pick} the best model.

Our initial scoring approach for the data challenge was a linear combination of various metrics (\S \ref{ssec:evalrs_scoring}), whereby the corresponding weights per metric were chosen on the basis of our experience and from the data obtained from the first stage of the challenge. Distilling our learnings, we identify two key properties/capabilities which a good scoring function should possess:
\begin{enumerate}
    \item \textbf{Metric Importance (P1)}: Given the non-determinism of model selection in the multi-objective setting, practitioners \textit{must} make operational decisions regarding metric importance (e.g. Accuracy is 2x more important than Fairness). An ideal scoring function should thus allow the relative importance of each metric to be specified precisely.
    
    \item \textbf{Scalability (P2)}: If we desire to accurately estimate the true Pareto Frontier in $\mathbb{R}^N$, it would thus require $O(2^N)$ data points (\S \ref{ssec:background_moe}). An ideal scoring function should thus be scalable with respect to the number of metrics involved, i.e., it should not require an unreasonable magnitude of data (e.g. exponential) in order to give reasonable results.
\end{enumerate}

While the approach used in \S\ref{ssec:evalrs_scoring} respected \textbf{P2}, it struggled with \textbf{P1}. In particular, it confounded metric aggregation and metric importance: since different metrics do not measure the same quantities and might not have the same scale and distribution, they should \textit{not} be directly compared. For example, Hit-Rate measures retrieval accuracy and its values might fall into the range ($10e^{-2}, 10e^{-1})$ whereas MRED is a measure of fairness and its values might fall in the range ($10e^{-3}, 10e^{-2})$. It is not immediately obvious what the appropriate scaling factor is, if it is even linear, and whether a summation of both terms can be interpreted meaningfully. Our scoring approach in EvalRS falls short as it did not clearly distinguish incommensurability and importance, resulting in a less than optimal weighting scheme (see \S\ref{ssec:analysis}).

\subsection{Rebooting EvalRS 2022: Evaluating new approaches through back-testing}
\label{ssec:approach}

Considering the properties of an ideal scoring function, we propose a more principled approach toward \textit{model selection} in the paradigm of rounded evaluation. While multi-objective model selection cannot have a \textit{determinate, single} answer, we put forward a motivated and systematic approach to satisfy our \textit{desiderata}. Our main insight is that only by first tackling the problem of \textit{incommensurability}, can we then appropriately address \textbf{P1}. Our proposed approach hinges on this, and learns the optimal-trade off between \textit{pairs} of metrics based on available data to maintain an $O(N)$ complexity (\textbf{P2}).


\begin{figure}[h!]
\centering
\includegraphics[width=0.32\linewidth]{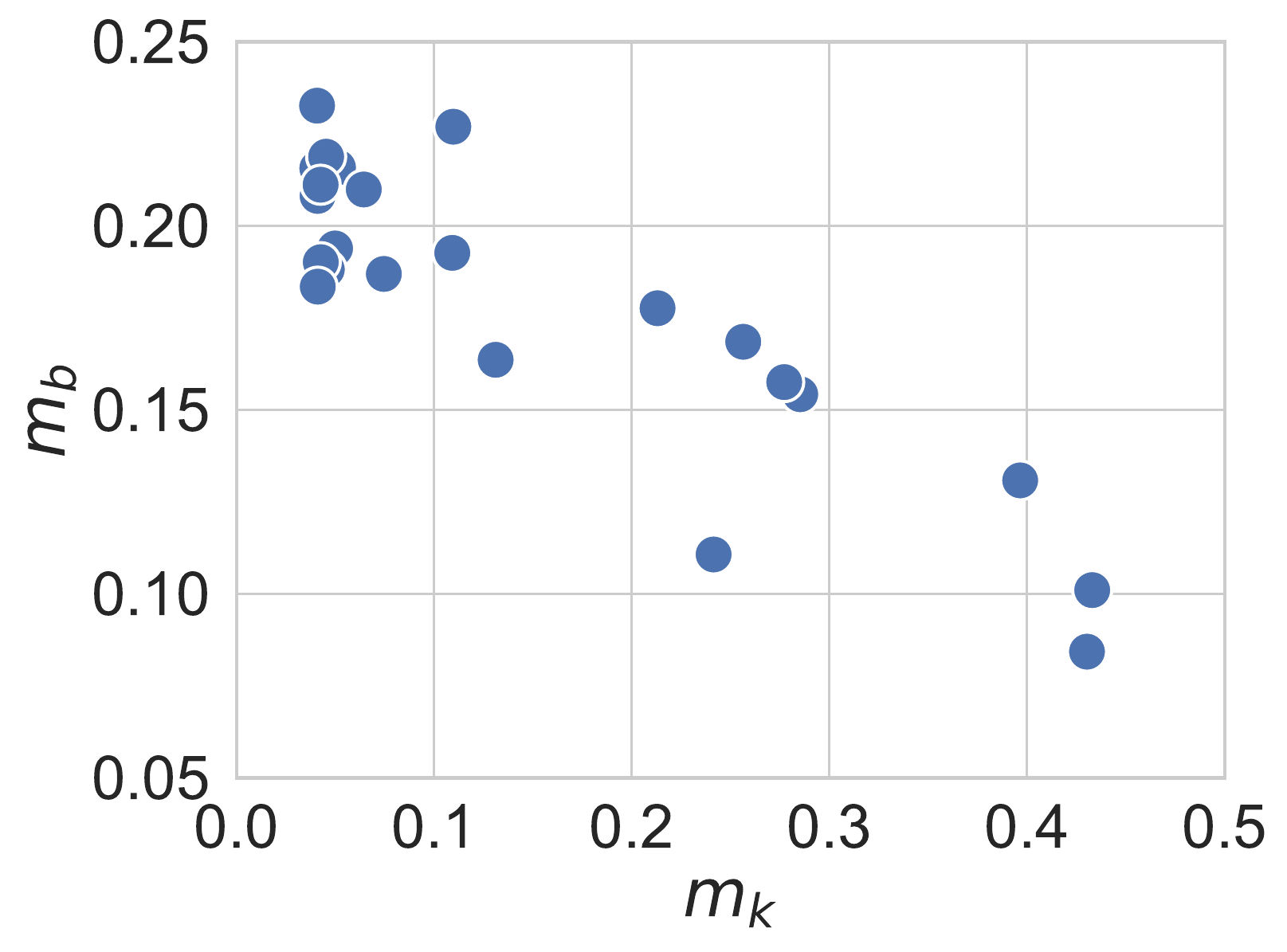}
\includegraphics[width=0.32\linewidth]{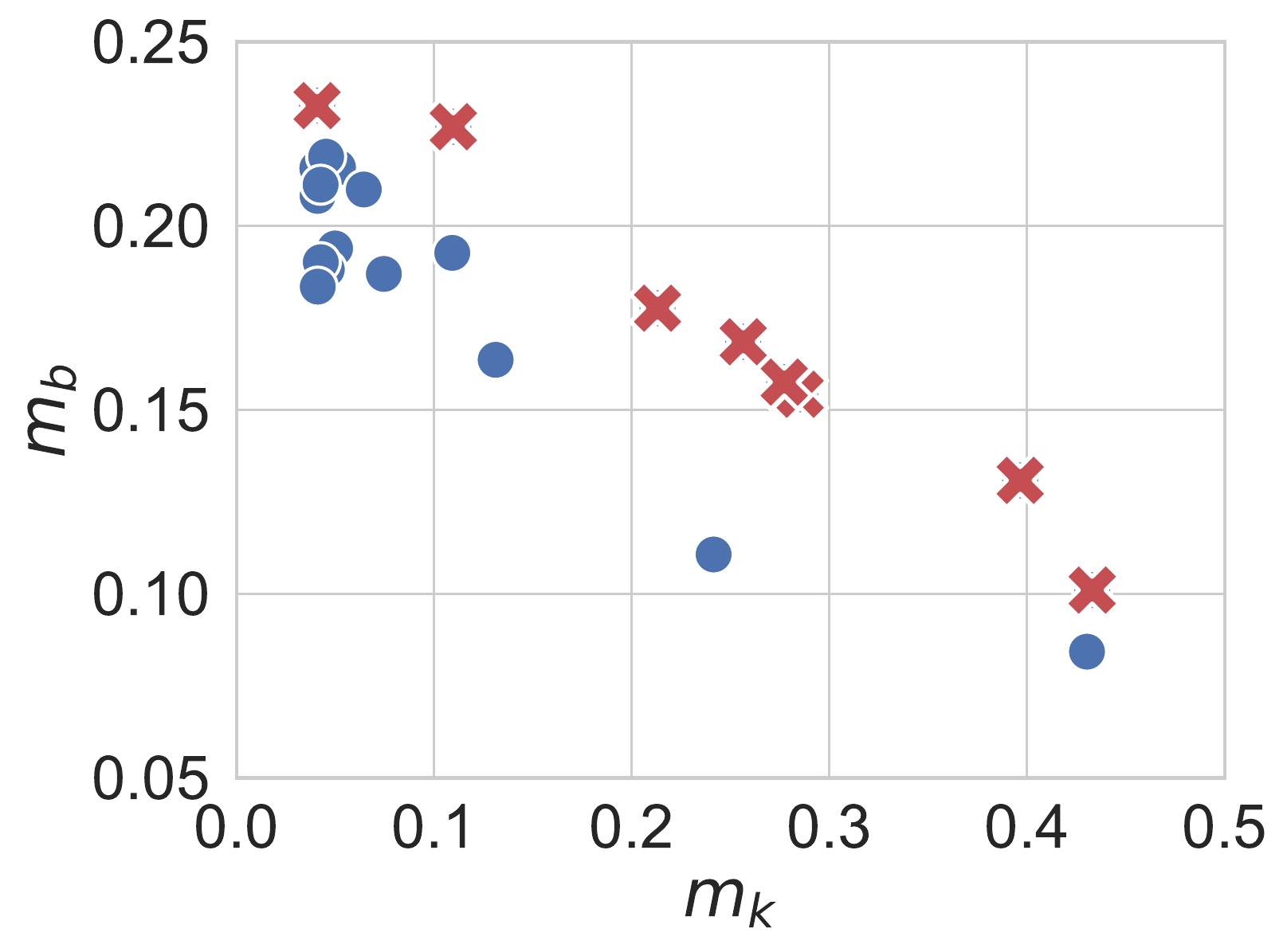}
\includegraphics[width=0.32\linewidth]{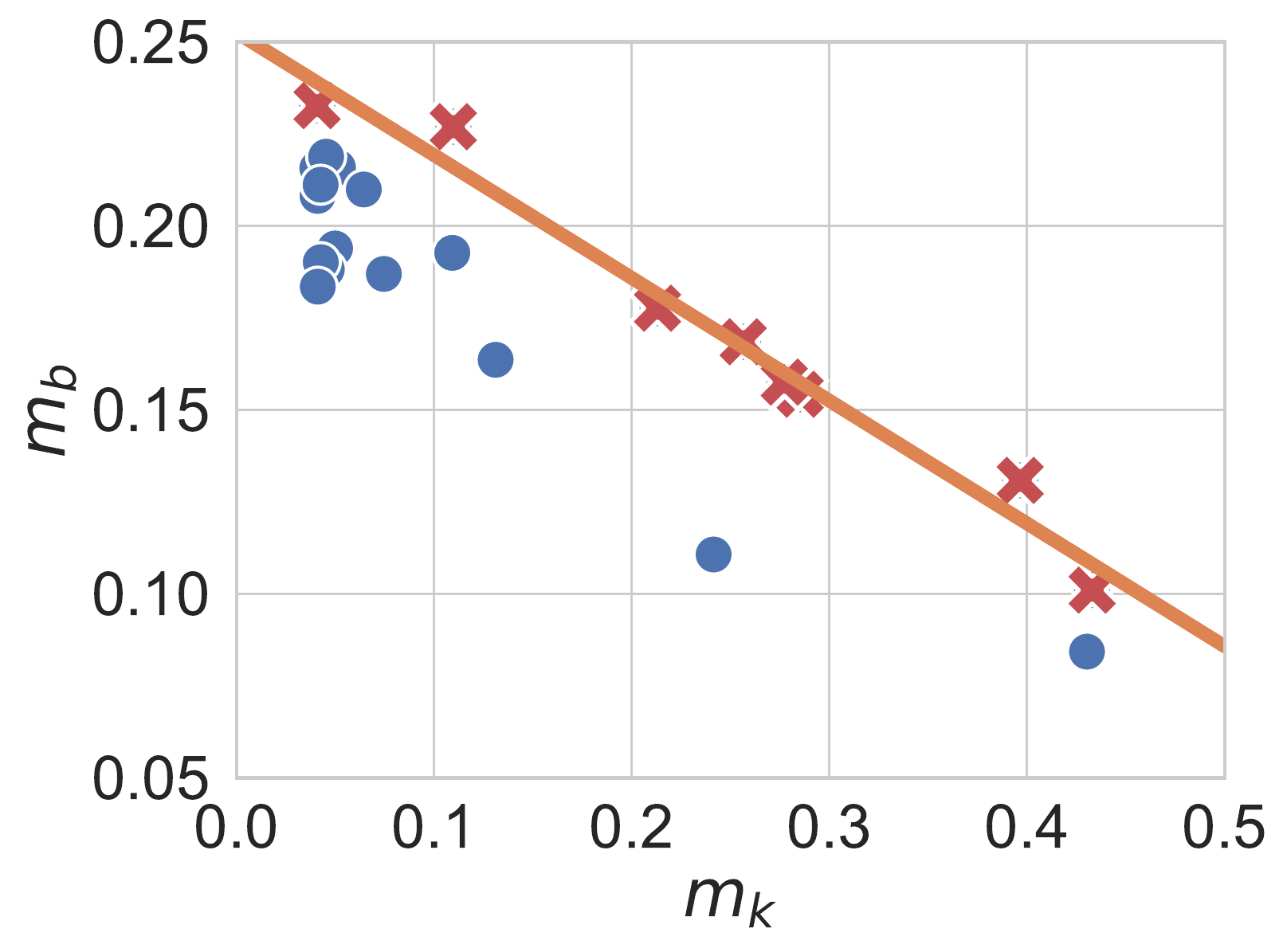}
\caption{From left to right: (1) $<m_b,m_k>$ data points are obtained via competition submissions or simulation. (2) Pareto non-comparable points are extracted from data. (3) Linear regression performed on Pareto non-comparable points to obtain a best estimate of the optimal trade-off curve.}
\label{fig:ev_estimation}
\end{figure}

\subsubsection{Definitions}
We introduce notation here, which will be expanded upon in subsequent sections. Let $\{m_0, .., m_{k-1}\}$ be a set of $k$ metrics used to assess a model $M$, and let $m_b \in \{m_0, .., m_{k-1}\}$ be a base metric used as a base unit. Subsequently, let $EV_b^i(\cdot)$ be a function that for a given metric value of $m_k$, returns the value of $m_b$ an optimal model should have. 

\subsubsection{Design Choices}
\begin{enumerate}
    \item Base Metric ($m_b$): We select \textit{Hit Rate} as $m_b$ given that model accuracy, though imperfect, is strongly correlated with model performance. This provides a common and familiar unit to reason about performance. In addition, we believe there is a priority on being accurate first, at least to some degree, before considering other performance metrics.
    \item Optimal Trade-off Function ($EV_b^i(\cdot)$): We obtain $EV_b^i(\cdot)$ by learning it from data\footnote{We assume here that the relationship between $m_b$ and $m_i$ is unknown or merely anecdotal. If the true functional relationship between $m_b$ and $m_i$ is known, then it is preferred over estimating with data.}. Referring to Fig. \ref{fig:ev_estimation}, given multiple data points $D_b^i = \{<m_b^0, m_i^0>, ..., <m_b^n, m_i^n>\}$, obtained either via simulation --- e.g., by learning a family of models across different objective weighting schemes \cite{Dosovitskiy2020You} --- or leaderboard submissions, we first extract the Pareto non-dominated points \cite{paretoset} from the set of data points. Then, we fit a curve to said Pareto non-dominated points in order to obtain the best-guess\footnote{It is not know whether our data contains the true Pareto Optimal points} optimal trade-off between $m_b$ and $m_i$. The choice of function to perform the curve fit is a design choice and for the rest of this section we restrict ourselves to a linear relationship.

    \item{Scoring Function ($S_p(.)$)}: We define the scoring function as follows,
        \begin{equation}
        \begin{aligned}
        \textrm{Given }m_b \in \{m_0,..,m_{k-1}\} &\textrm{ and } \{w_0,... w_{k-1}\}\setminus w_b \\
        &\textrm{where } w_i \in (0,1) \\
        S_p(m_0, ..., m_{k-1})&= \sum_{i=0, i \neq b}^{k-1} (1-w_i) \cdot m_b-w_i \cdot EV_b^i(m_i)\\
        &=\sum_{i=0, i \neq b}^{k-1} \Delta_i
        \end{aligned}
        \end{equation}
        
Qualitatively, we convert metric $m_i$ into the units of $m_b$ via $EV_b^i(.)$ and subsequently, given an importance ratio $w_i$, we obtain the performance differential $\Delta_i$, which represents whether the current model M in consideration is underperforming ($\Delta_i < 0$) , over performing ($\Delta_i > 0$) or at parity ($\Delta_i = 0$) with respect to its expected optimal performance given $m_i$ and importance weight $w_i$. Finally, we take a summation of $\Delta_i$ for each given $<m_i, m_b>$ pair to obtain a single score.
\end{enumerate}

\subsection{Analysis and Interpretation}
\label{ssec:analysis}

We provide further intuition behind our proposed scoring method by making an analytical comparison with the scoring method used in EvalRS 2022, which reveals critical insights about the underlying mechanism of our new approach.

\subsubsection{Analytical Comparison}

Given the linearity assumptions between $m_b$ and $m_k$, we can make an analytical comparison between our original scoring approach, henceforth $S_o$, and the proposed approach, $S_p$. As we will see, both are essentially linear transformations of their inputs (i.e. $<m_0, ..., m_{k-1}>$), with the difference lying in how their coefficients are derived. We can therefore rewrite the former scoring approach into the same form as the latter to compare said coefficients. For the sake of analysis, we restrict ourselves to two dimensions, i.e.,  we focus on ~\texttt{Hit Rate} and \texttt{MRED (User Activity)} and drop unnecessary notation.

In the proposed approach, the scoring function can be written as,

\begin{equation*}
\begin{aligned}
S_p(hr, mr_{ua}) &= (1- w) \cdot hr - w \cdot EV(mr_{ua})\\
                   &= (1- w) \cdot hr - (w \cdot c_1 \cdot mr_{ua} + w \cdot c_2)\\
                   &= (1- w) \cdot hr - w \cdot c_1 \cdot mr_{ua} - w \cdot c_2
\end{aligned}
\end{equation*}
Where $c_1$ and $c_2$ are the coefficient and intercept recovered from linear regression, $w$ is the importance ratio between $mr_{ua}$ and $hr$. Similarly, the scoring function from EvalRS 2022 (see Eqn. \ref{eqn:evalrs_norm}) can be re-written as,

\begin{equation*}
\begin{aligned}
S_o(hr, mr_{ua}) &= k_1 \cdot \bar{hr} + k_2 \cdot \bar{mr}_{ua}\\
                 &= k_1 \cdot \frac{hr-h_2}{h_1-h_2} + k_2 \cdot \frac{mr_{ua}-mr_2}{mr_1-mr_2}\\   
                 &= \frac{k_1 \cdot hr}{h_1-h_2} - \frac{k_1 \cdot h_2}{h_1-h_2} + \frac{ k_2 \cdot mr_{ua}}{mr_1-mr_2} - \frac{k_2 \cdot mr_2}{mr_1-mr_2}   
\end{aligned}
\end{equation*}

If we let $\alpha = \frac{k_1}{h_1-h_2}$ and $\beta = \frac{k_2}{mr_1-mr_2}$ then,

\begin{equation*}
\begin{aligned}
S_o(hr, mr_{ua}) &= \alpha \cdot hr - \alpha \cdot h2 + \beta \cdot mr_{ua} - \beta \cdot mr_{2}\\
                 &\propto (1-w) \cdot hr + (1-w) \cdot\frac{\beta}{\alpha} \cdot mr_{ua} - (1-w) \cdot (h_2 +  \frac{\beta}{\alpha} \cdot mr_{2})
\end{aligned}
\end{equation*}

Comparing coefficients, we see that,

\begin{equation}
\label{eqn:coeff}
\begin{aligned}
 w \cdot c_1 = -(1-w) \cdot\frac{\beta}{\alpha}
\end{aligned}
\end{equation}

and $w \cdot c_2 = (1-w) \cdot(h_2 + \frac{\beta}{\alpha} \cdot mr_{2})$\footnote{Note that we can discard $c_2$ for the sake of rank analysis as it only impacts absolute score and not relative score.}. By reintroducing $\alpha$ and $\beta$ terms, we obtain,

\begin{equation*}
\begin{aligned}
 -(1-w) \cdot \frac{\beta}{\alpha} &= -(1-w) \cdot\frac{k_2}{k_1} \cdot \frac{h_1-h_2}{mr_1-mr_2}
\end{aligned}
\end{equation*}

which in itself has the canonical form of a gradient/slope between two points in $\mathbb{R}^2$, with scaling factors. In other words, the original scoring mechanism attempts to estimate the gradient of the optimal trade-off curve using $<h_1, m_1>$ and $<h_2, m_2>$, i.e., the ``best model'' in the data and a \textit{baseline} model both determined by us. In addition, the ratio $-(1-w) \cdot \frac{k_2}{k_1}$ can be interpreted as our ratio $w$. The negation comes from the fact that our proposed scoring approach takes a differential between the two metrics rather summation.

\subsubsection{Interpretation}

We compare the coefficients obtained in our proposed approach with the ones used during the data challenge. Given that there are two degrees of freedom, $w$ and $c_1$, we fix $w=0.5$ to allow for an unbiased interpretation of $c_1$\footnote{$w$ and $1-w$ cancel each other when $w=0.5$.}.

The first major difference is the negation term on the RHS of Eqn. \ref{eqn:coeff}, highlighting that the original scoring mechanism did not respect the principles of Pareto Optimality, i.e., assuming equal metric importance,  models which lie on the Pareto Optimal front will \textit{not} be assigned an equal score.

We now compare the gradient obtained via normalization, $c_1^o=\frac{h_1-h_2}{mr_1-mr_2}$, and regression, $c_1^p$. The value of $c_1^o$ used (implicitly) during EvalRS is given as $44.399$ whereas the $c_1^p$ we obtain is $-7.944$. Ignoring magnitude differences, a difference in sign implies that the models used to estimate $c_1^o$ did not reflect the competing nature of the two metrics involved. We observe a similar trend in other <\texttt{Hit Rate}, \texttt{MRED}> pairs. Incidentally, this negated the sign error we highlighted in Eqn. \ref{eqn:coeff}, allowing us to estimate\footnote{Given the differences in the scoring mechanism, and constraints imposed by the proposed one, an exact one-to-one comparison is not possible.} and compare the metric importance implied during EvalRS 2022 --- we find that compared to what the data says, \texttt{MRED (User Activity)} was effectively given $14$x more importance than \texttt{Hit Rate}, much more than we had intended.

Drawing from the link between the old and the new, the proposed approach can thus be seen as a form of data-driven, optimal trade-off, normalization with knobs for controlling metric importance --- a consequence of (1) converting metrics into a single base metric and (2) learning the optimal trade-off function from data. As compared to $S_o$, our approach more faithfully respects the relationship between metrics, evident in how we approximate the gradient $c_1$ using multiple data points as opposed to $S_o$, which effectively utilizes only two points: the current ``best'' model and a baseline, which introduces significant noise due to human choice. It also more faithfully respects the principles of Pareto Optimality when considering a \textit{pair} of metrics: if we assign equal importance to a pair of metrics ($w_i=0.5$), our approach assigns approximately\footnote{Noise is introduced by regression.} equal scores to models that adhere to the optimal trade-off, something which $S_o$ does not guarantee. As a result, we are able to introduce importance weights $w_i$ that more accurately encode the trade-off we desire. Collectively, our proposed scoring approach provides a more systematic lens toward scoring in the rounded evaluation paradigm.

\subsection{EvalRS Post-Mortem}

\begin{figure}[t!]
\centering
\includegraphics[width=0.41\linewidth]{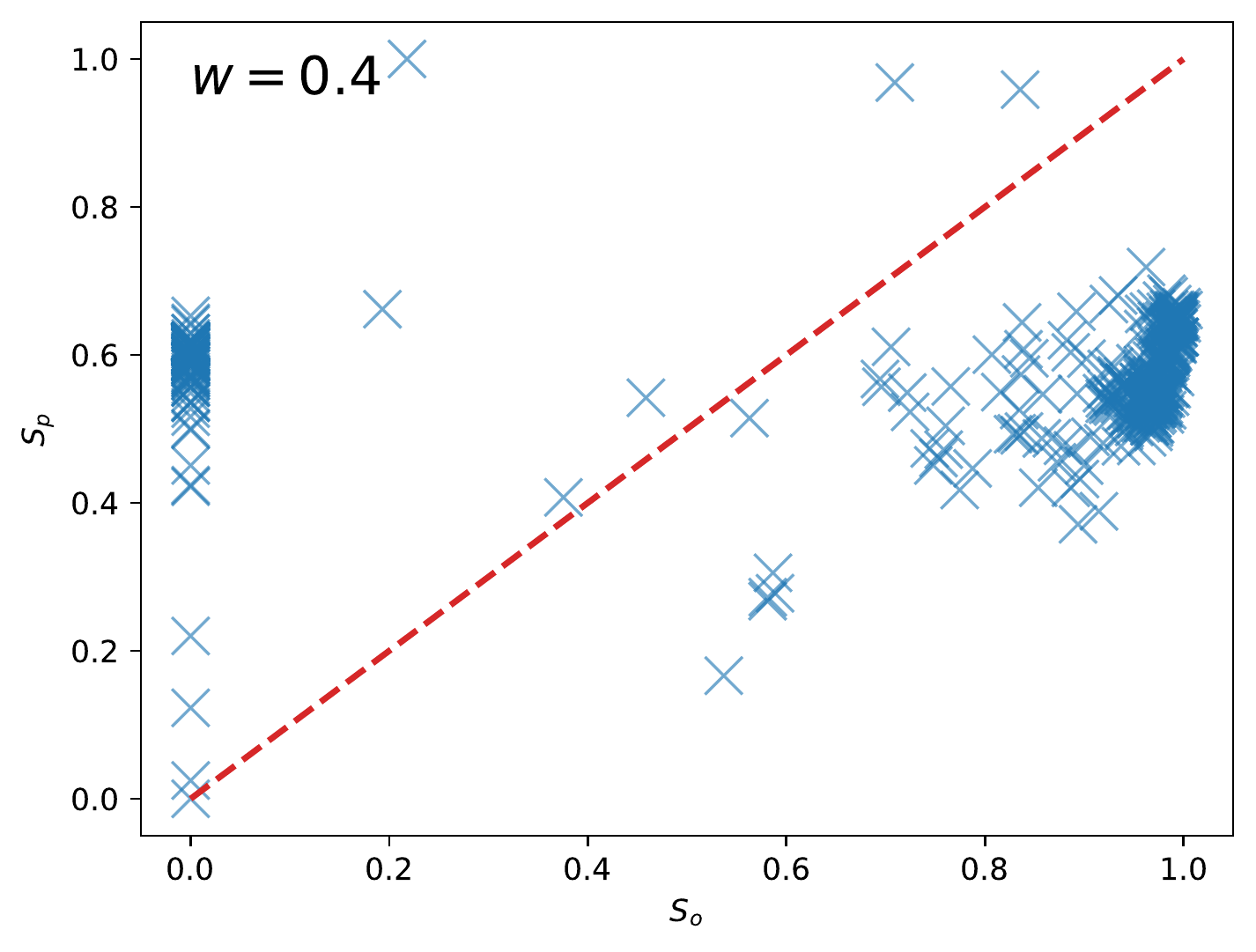}
\includegraphics[width=0.41\linewidth]{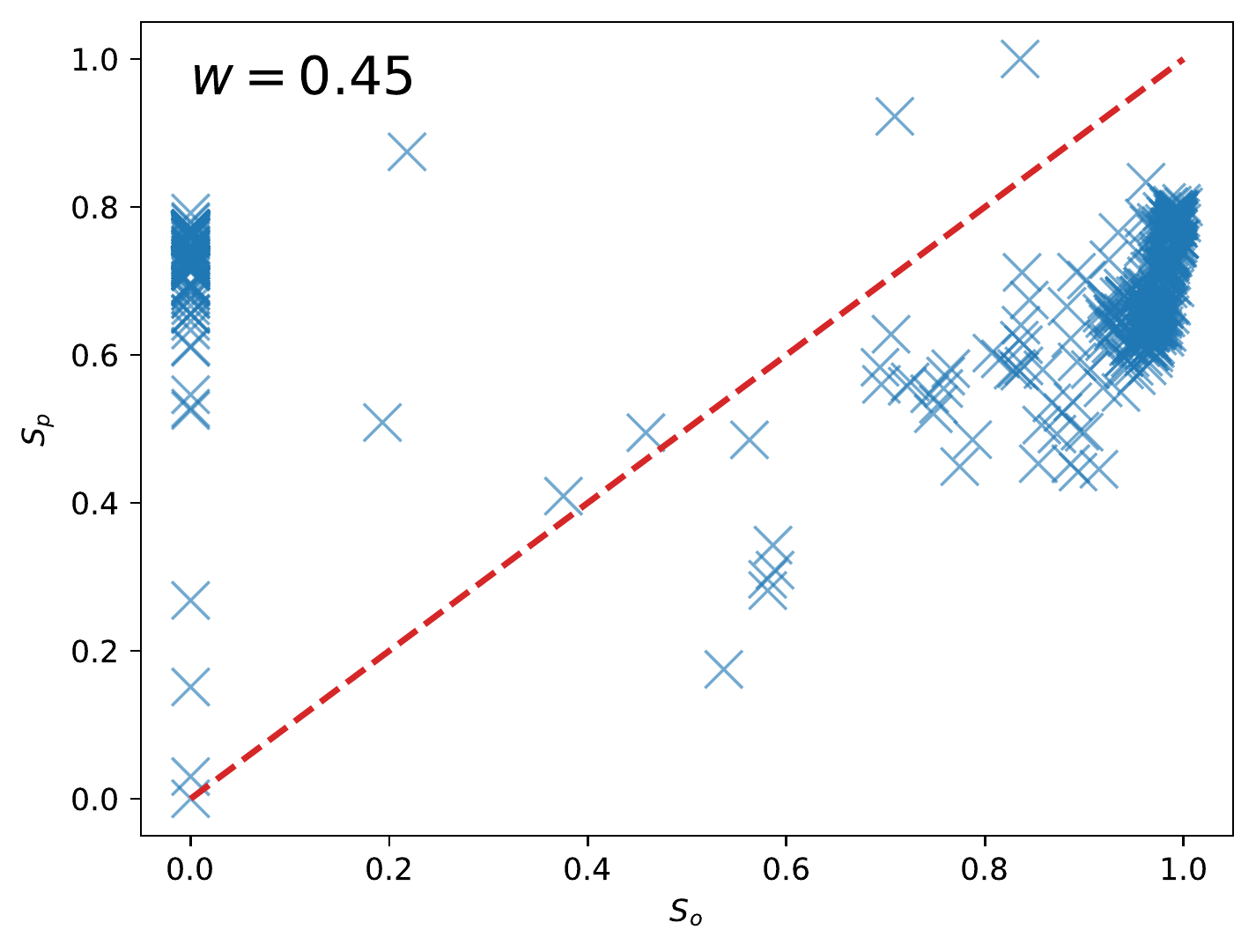}
\includegraphics[width=0.41\linewidth]{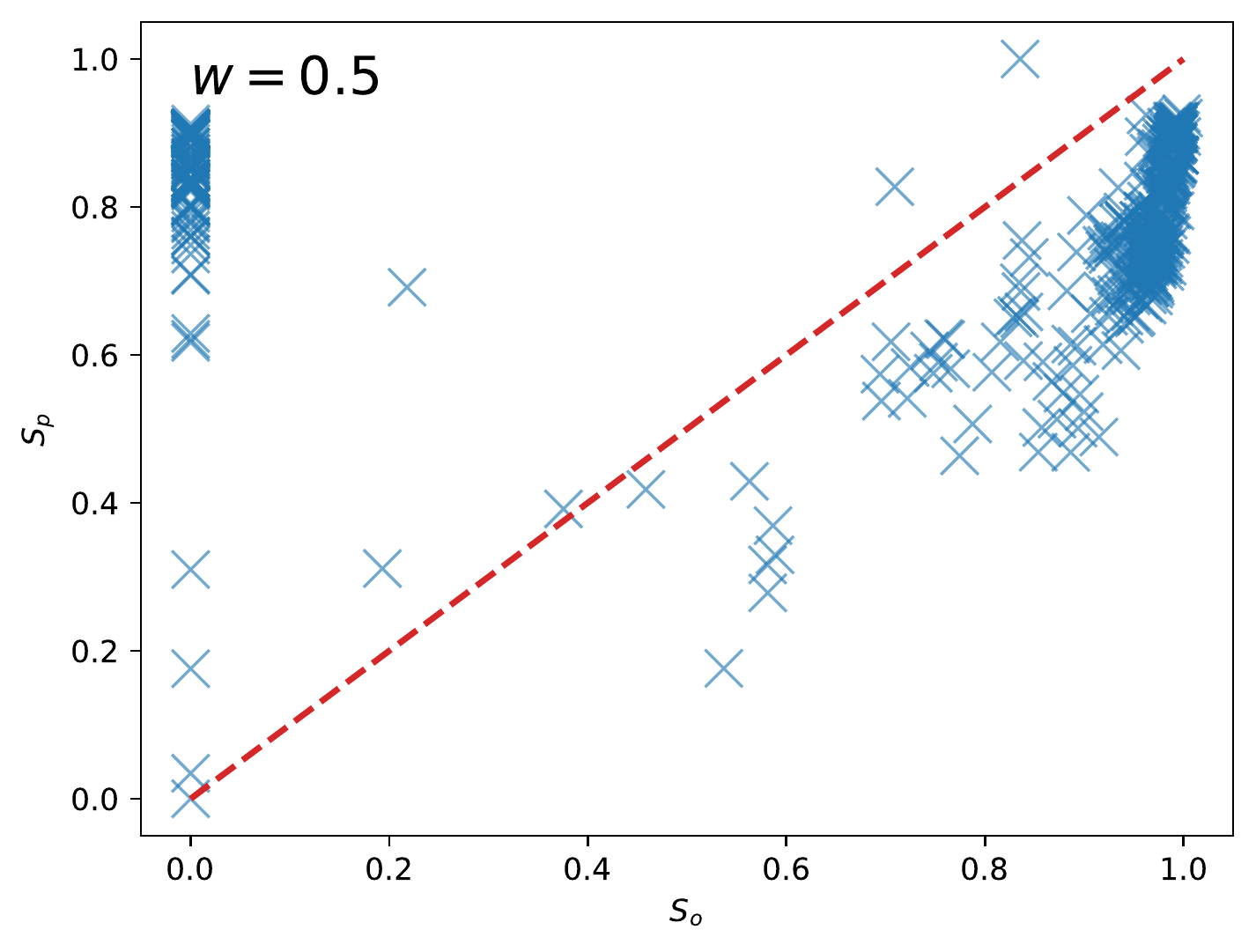}
\includegraphics[width=0.41\linewidth]{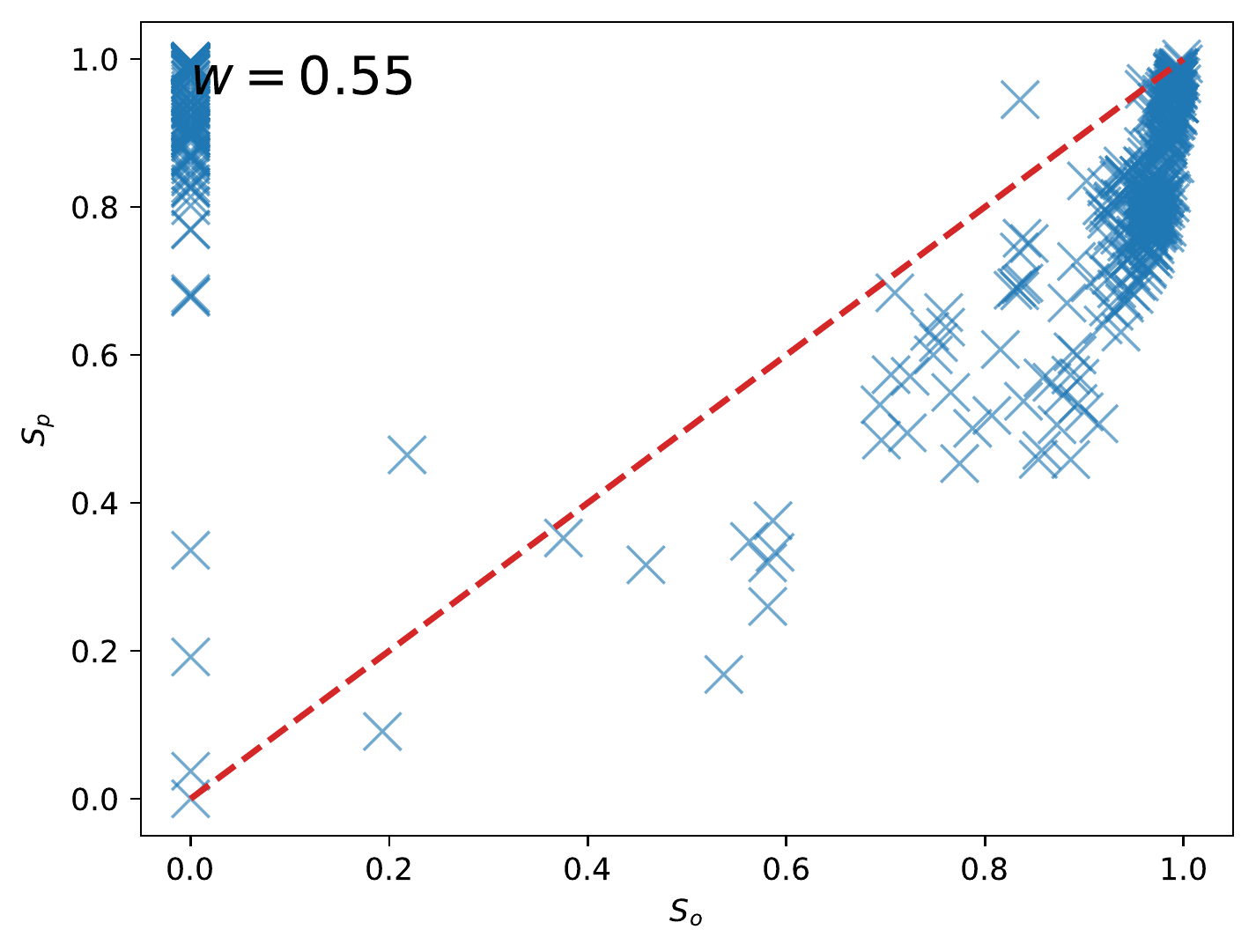}
\caption{Scoring of submissions based on our proposed scoring $S_p$ plotted against the original EvalRS scoring $S_o.$ for various settings of $w$. Red-dotted line denotes $y=x$.}
\label{fig:rescoring_naive_vs_new}
\end{figure}

Circling back to where we began, we apply our proposed approach to the competition data obtained from EvalRS 2022, allowing us to better understand how submissions were ranked during the challenge, whilst elucidating the merits of our proposed approach. Our analysis refers to Fig. ~\ref{fig:rescoring_naive_vs_new} which visualizes $S_p$ against $S_o$, normalized to a range of $(0, 1)$, for various settings of $w$, whereby $w$ is uniform for all $\{w_0,..,w_{k-1}\}\setminus w_b$. Note that models assigned $S_o = 0$ were due to a minimum \texttt{Hit Rate} threshold requirement as part of the scoring.

Setting $w = 0.5$, i.e., giving accuracy and fairness equal importance, our first observation is that the EvalRS scoring was effectively giving more weight to fairness than toward accuracy. This is evident in how most of the EvalRS scores sit below the $y=x$ line (red, dashed). We observe even greater correlation as $w$ is increased to $0.55$ (bottom right) and beyond (not shown). 

We also observe that many of the models assigned a score of $S_o = 0$ are given are high ranking in $S_p$. Indeed, one of the motivations when assigning a minimum threshold was to avoid assigning high scores to models that had too low an accuracy\footnote{It is easier to score in fairness with low accuracy.}. While this may have worked practically for EvalRS, we argue that such an approach promotes an unintentional ``gaming'' of the leaderboard --- many submissions had a \texttt{Hit-Rate} which were slightly above the minimum threshold (Median \texttt{Hit-Rate} = 0.0175) which, in hindsight, restricted the spirit of the challenge. 

We believe that our newly proposed approach can offer some resilience towards ``gaming'', and be especially helpful for future events: by updating $EV(.)$ iteratively with each new submission, the notion of optimal is constantly updated rather being chosen \textit{a priori}. This incentivizes competition, and circumvents a common pitfall in evaluation as summarized by Goodhart's law ---  ``When a measure becomes a target, it ceases to be a good measure''.

\section{The Future: Guidelines for Future Challenges}
\label{sec:guidelines}

We illustrate in this section our set of minimal guidelines that we believe can make up for a successful holistic evaluation of recommender systems and challenges. We divide the guidelines in two: the first part is a more general guideline on the organization of the challenges while the second part are suggestions on how to design a scoring mechanism .

\subsection{Guideline: Competition Structure}
 
\subsubsection{Dataset}
With ever increasing modelling capabilities \cite{NEURIPS2020_1457c0d6, Chowdhery2022PaLMSL} and growing sizes of datasets \cite{Zheng_2022_CVPR}, the compute requirements to participate in data challenges has risen accordingly, with the problem further exacerbated when multiple objectives are in concern. By preparing a smaller yet signal-dense dataset, the data challenge is made more accessible, encouraging contributions from participants of varying backgrounds. Another critical requirement for running an multi-objective evaluation data challenge is rich metadata, which enables slicing of data into interesting facets, and may unlock more in-depth evaluation approaches --- e.g., given song lyrics, we could assess song similarity in another way.

\subsubsection{Evaluation Metrics}

Beyond the obvious contenders for accuracy metrics (Hit-Rate, MRR, NDCG), and beyond-accuracy metrics (diversity, serendipity, population bias, fairness), we believe that further innovation into beyond-accuracy metrics is critical to the adoption and utility of multi-objective evaluation. As a starting point, ``being less wrong''\cite{DBLP:journals/corr/abs-2111-09963} metrics introduced in EvalRS is fertile ground upon which more innovative metrics can be developed -- for example, Large Language Models could be used to elicit ``similarity judgements'' \cite{Rosenbaum2022} to complement the representational approach. EvalRS also highlighted that improved formulations of classical metrics, such as fairness \cite{10.1145/3471158.3472260}, are also opportunities for improvement.

\subsubsection{Evaluation Structure}
A successful challenge depends also on dealing with practical considerations:

\begin{enumerate}
\item{Prevent Leaderboard Hacking}: It is important to find ways to prevent users to game the challenge (e.g. with a single test set,  estimating test parameters may be feasible). Bootstrapped cross-validation methodologies makes it harder to game the leaderboard and allow to also report standard deviation over the results.

\item{Evaluation Platform}: We suggest the adoption of an open source tool, such as RecList \cite{DBLP:journals/corr/abs-2111-09963}, in such a way that progress can be easily made public and that the challenge can be replicated in the future. Moreover, participants can evaluate their models offline (long after the event is concluded) without the need to get third-party access to an external platform.

\end{enumerate}

\subsection{Guideline: Scoring Methodology}

We conclude by providing guidelines on how to best use our approach, and which choices are left to be made.

\subsubsection{Base Metric}

As the base metric we suggest an accuracy-based metric (e.g. \texttt{Hit Rate}), since we believe that accuracy is easy to understand and still a very important component (even if not the only one).
    
\subsubsection{Optimal Trade-Off Function} 

The choice of the family of functions that are used to learn the optimal trade-off function from the data is critical to the success of this approach and will depend on the relationship between the metrics observed after Phase 1 or based on expert knowledge.

\subsubsection{Estimation from Data}

The two-stage approach championed by EvalRS 2022 is an effective paradigm: easy to understand by practitioners, and convenient to collect the data we need to estimate the optimal trade-off between metrics. After the end of Phase 1, the data can be used for the initial estimation, and be updated with newer submissions from Phase 2.

\subsubsection{Weighting Scheme}
While algorithmic approaches can inform this decision this choice is ultimately an operational decision and should reflect the objectives of the challenge. We believe this to be an important consequence of our formulation; indeed, a strength of our proposal is the clean separation between what should be left to the human and what should be inferred from the data.

\section{Conclusion}

In order to develop recommender systems with more predictable ``in-the-wild'' performance, it is necessary to evaluate models with measures other than accuracy; that is, beyond-accuracy measures must become defaults in the modern-day evaluation toolkit. EvalRS 2022 was our first attempt at fostering awareness of the rounded evaluation of recommender systems. It introduced practical and reusable code for the community. We leverage the experience we garnered from our unique data challenge to further refine definitions and requirements for multi-objective evaluation and to formulate a principled framework and a set of guidelines for performing multi-objective evaluation.

As we prepare the second edition of EvalRS, we released the dataset and the code used to prepare this contribution \footnote{Available at \url{https://github.com/RecList/e-pluribus-unum-evalrs-2022}}, hoping to benefit not just other organizers, but the recommender system community at large. In particular, even if data challenges are rarely completely representative of real-world deployments, we do believe industry practitioners are eager to incorporate more rounded evaluation in their pipelines \cite{natureMIchallenge}: when they do, they will indeed face the problem of selecting which architecture, among many, should be promoted to production. While hardly the last word on the topic, we hope these reflections will help them take the first few steps more firmly.

\section*{Limitations}

The contributions we have presented in this work come with some limitations. First of all, we make linear assumptions when defining a scoring function, this can strongly affect how models behave if relations between the metrics are non-linear. In addition, since we perform pairwise comparisons between the base metric and other metrics, we do not model the effects between the non-accuracy metrics. This also means that our approach may not fully respect pareto optimality beyond two dimensions (it is worth noting that this is a design choice to combat combinatorial explosion, but is still a limitation).

Our proposed guidelines for challenges come with some inherent limitations. We assume, for example, that all challenges can be split into two phases, and this might not be true for fast-paced challenges. In addition to this, we assume the test data is freely available to participants; while we used bootstrapping to reduce the possibility of test-set leakage/leaderboard hacking, this cannot completely mitigate the issue, and many organizers might prefer not to share the test set.

\begin{acks}
FB is supported by the Hoffman–Yee Research Grants Program and the Stanford Institute for Human-Centered Artificial Intelligence. GA is supported by Fondazione Cariplo (grant No. 2020-4288, MONICA).

We wish to thank Dietmar Jannach for helpful comments on a previous draft of this paper. We wish to also thank all the participants of EvalRS 2022 -- starting from Tobias Schnabel --, and the organizers of CIKM -- starting from Surya Kallumadi -- for providing a fantastic venue and great support: this work would have been impossible without such a successful event. No manager and no Italian bureaucrat were harmed in the course of running our experiments.

\end{acks}

\bibliographystyle{ACM-Reference-Format}
\bibliography{acmart,anthology,cikm}





\end{document}